\documentstyle[12pt,amssym,aas2pp4]{article}
\def\Msolar{{M$_{\odot}$\,}}
\def\arcsec{{$^{\prime\prime}$}\,}
\newcommand{\secpoint}{\mbox{$''\mskip-7.6mu.\,$}}
\newcommand{\minpoint}{\mbox{$'\mskip-5.6mu.\,$}}

\received{Jun 23, 1998 } 
\accepted{Sep 14, 1998 }

\slugcomment{To appear in ``The Astronomical Journal''}

\begin{document}

\title{THE LUMINOSITY FUNCTION OF OMEGA CENTAURI\altaffilmark{1}}

\altaffiltext{1}{Based on observations with the NASA/ESA {\it Hubble
Space Telescope}, obtained at the Space Telescope Science Institute,
which is operated by AURA, Inc., under NASA contract NAS5-26555}

\author{Guido De Marchi} \affil{European Southern Observatory,
Karl-Schwarzschild Strasse 2, D--85748 Garching, Germany\\
{\rm \vspace*{0.1in} To appear in ``The Astronomical Journal'' \\
 Received Jun 23, 1998 / Accepted Sep 14, 1998}}
\authoremail{demarchi@eso.org}

\begin{abstract}

Deep {\it HST}--WFPC2 observations of the stellar population in the
globular cluster $\omega$\,Cen (NGC\,5139) have been used to measure
the luminosity function of main sequence stars at the low-mass end. Two
fields have been investigated, located respectively $\sim 4\minpoint6$
and $\sim 7^\prime$ away from the cluster center, or near the
half-light radius of this cluster ($r_{hl} \simeq 4\minpoint8$). The
color-magnitude diagrams derived from these data show the cluster main
sequence extending to the detection limit at $I \simeq 24$. Information
on both color and magnitude is used to build the luminosity functions
of main sequence stars in these fields and the two independent
determinations are found to agree very well with each other within the
observational uncertainty. Both functions show a peak in the stellar
distribution around $M_I \simeq 9$ followed by a drop at fainter
magnitudes well before photometric incompleteness becomes significant,
as is typical of other globular clusters observed with the {\it
HST}. This result is at variance with previous claims that the
luminosity function of $\omega$\,Cen stays flat at low masses, but is
in excellent agreement with recent near-IR NICMOS observations of the
same cluster.

\end{abstract}


\section{Introduction}

With a total mass of $\sim 3.9 \,\, 10^6$\,\Msolar (Meylan 1987),
$\omega$\,Cen (NGC\,5139) is the most massive globular
cluster (GC) in the Galaxy. Such a large mass, coupled with the strong
anisotropy measured in the velocity dispersion (Meylan 1987)
implies that, except probably in the innermost regions, the cluster
has not yet reached relaxation through energy equipartition and,
therefore, the mass function (MF) of main sequence (MS) stars should
still reflect the unperturbed IMF, as no dynamical modifications should
have occurred. It is, thus, somewhat surprising that Elson et al.
(1995) obtain for $\omega$\,Cen a luminosity function (LF) which
deviates substantially at the low mass end from that of other clusters
with similar metal content. In fact, while low metallicity clusters
such as NGC\,6397 (Paresce, De Marchi, \& Romaniello 1995; Cool,
Piotto, \& King 1996), NGC\,7078 (De Marchi \& Paresce 1995), NGC\,7099
(Piotto, Cool, \& King 1997),  NGC\,6254 and NGC\,6809 (De Marchi \&
Paresce 1996), NGC\,6656 (De Marchi \& Paresce 1997) all show a LF that
reaches a peak at $M_I \simeq 9$ and then drops all the way to the
detection limit, the results published by Elson et al. for
$\omega$\,Cen feature a flat LF below $M_I\simeq 8$, with no signs of a
drop.

The actual shape of the LF may vary with the location inside the
cluster because of dynamical evolution, both internal due to two-body
relaxation and external due to the interaction with the Galactic tidal
field. Yet, when the LF is measured near the half-mass radius (i.e.
within 1-3 times $r_{hm}$) as for all the clusters above including
$\omega$\,Cen, and barring extreme cases of tidal stripping such as
those resulting from clusters venturing very close to the Galactic
bulge, one should expect only minimal deviations from the original
distribution of masses as De Marchi \& Paresce (1996, 1997) have shown
observationally, and as Richer et al.  (1991) had predicted.

If the results of Elson et al. were confirmed, however, they could
modify this picture. In fact, because $\omega$\,Cen is both too massive
to have undergone serious internal dynamical modifications in a Hubble
time, and is sufficiently far from the Galactic plane and bulge not to
have experienced major tidal stripping (see Dauphole et al. 1996), its
LF should reflect the IMF much more closely than any of the LFs of the
other clusters listed above. Since clusters with a similar metallicity
are expected to form with a similar IMF, the discrepancy between the LF
of $\omega$\,Cen and that of the other clusters would seem to suggest
that GCs are born with a steeply rising IMF that is substantially
modified in less massive clusters by dynamical evolution throughout the
clusters lifetime. Thus, in these cases the observed MF no longer
corresponds to the IMF.

Such a conclusion would call for a reinterpretation of most of the
recent HST-based studies of stellar populations in GCs and in the field
which today all point to a deficiency of low-mass stars in the IMF. The
implications that such a result could bear on the origin and nature of
dark matter in the Galaxy and the star formation mechanism are so
important that we must be able to exclude that the apparent anomaly of
the LF of $\omega$\,Cen reported by Elson et al. is not due to the data
reduction process or to the limitation of the data themselves.  Most of
Elson et al.'s conclusions are based on the analysis of a deep $I$ band
image, so that their LF is constructed using little or no color
information for MS stars. This is somewhat worrisome, as there are many
examples of GC LFs based on single-filter photometry (e.g. Fahlman et
al. 1989; Richer et al. 1991) which have later proved to be incorrect
at the faint end, although in these cases the error might have
originated mostly because of crowding.

To better understand this issue, I have used a larger set of WFPC2
images of $\omega$\,Cen now available taken from the HST archive, and
have subjected these data to exactly the same processing that I adopted
for all other clusters studied so far. In this paper, I describe the LF
obtained in this way, and compare it with those of other GCs observed
with the same instrumentation, as well as with the near-IR LF recently
obtained with NICMOS for the same object (Pulone et al. 1998).

\section{The Data}

\begin{figure}
\plotone{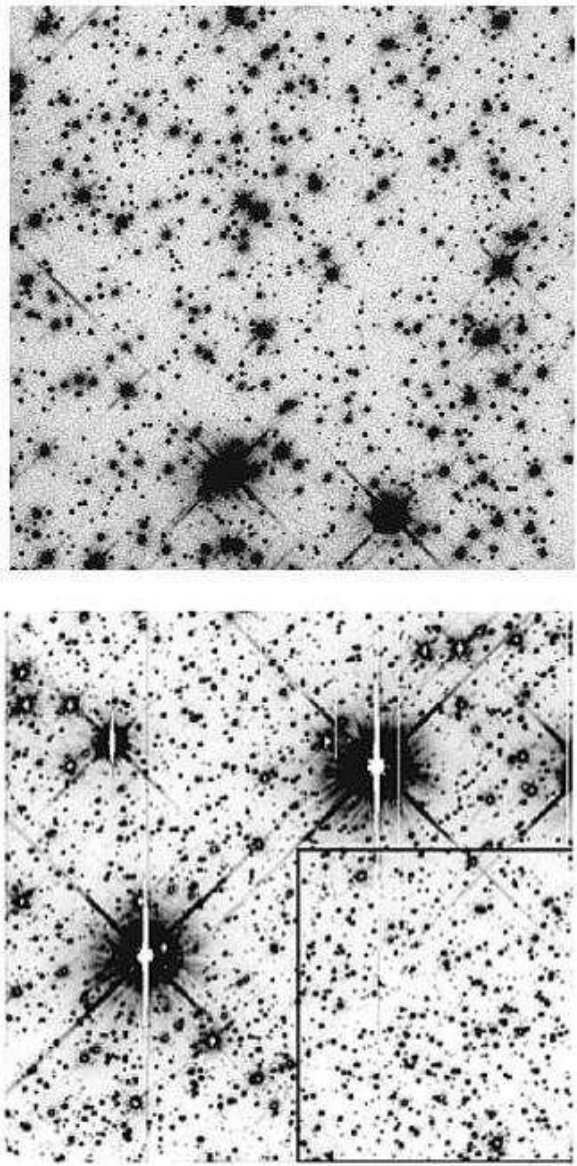}
\caption{Top: negative image of a region of $\sim 30$\arcsec $\times
30$\arcsec located $\sim 7^\prime$ SW of the center of $\omega$\,Cen as
seen through the F814W filter of the WFPC2 (PC chip) with an exposure
of 1800 s duration (Field\,1). Bottom: negative image of a region of
$\sim 30$\arcsec $\times 30$\arcsec located $\sim 4\minpoint6$ SE of
the center of $\omega$\,Cen observed with the F814W filter of the WFPC2
(PC chip) with an exposure of 4600 s duration (Field\,2); the box on
the lower right corner marks the region in which photometry was carried
out.}
\end{figure}

The data discussed in this paper have been obtained with the WFPC2 on
board the {\it HST}, using the F606W and F814W filters (Biretta 1996).
Two fields have been observed: the first, located $\sim 7^\prime$ SW of
the cluster center, was imaged on 1994 August 29 with a total exposure
time of 1800\,s in the F606W band and 2800\,s in the F814W filter,
while the second, located $\sim 4\minpoint6$ SE of the center, was
observed on 1995 June 4 with a total exposure time of 3000\,s in F606W
and  4600\,s in F814W, respectively. The images of the first field are
the same as those described in the Elson et al. paper (their Field\,2),
although they erroneously quote for it a distance of $17\minpoint4$
from the cluster center. The analysis described here has been limited
only to the PC chips in both fields, as the high stellar density makes
it difficult to obtain accurate photometry for faint stars in the WF
chips because of their larger plate-scale ($0\secpoint1$\,pix$^{-1}$ vs
$0\secpoint043$\,pix$^{-1}$), thus increasing the uncertainty on the
derived LF. The two PC fields used in this paper are shown in
Figure\,1.

After standard {\it HST} WFPC2 pipeline processing, including frame
registration and cosmic ray rejection, stellar fluxes have been
measured using the core aperture photometry technique (De Marchi et
al.  1993), with the same selection of parameters that I used in
previous investigations of other clusters with the same instrumentation
(see Paresce et al. 1995 and De Marchi \& Paresce 1995 for a detailed
description of our photometric analysis and for a discussion of the
current uncertainties in the reduction and calibration of WFPC2
photometric data).

With a detection threshold conservatively set at $5 \sigma$ above the
local average background in the PSF peak, I have selected a total of
1070 stars in Field\,1 with well defined fluxes in both bands. In
Field\,2, because of the presence of three very bright stars,
photometry has been restricted to the lower right portion of the image
(see Figure\,1), where I have found 830 objects satisfying the
detection criterion above in both filters.

\begin{figure}[h]
\plotone{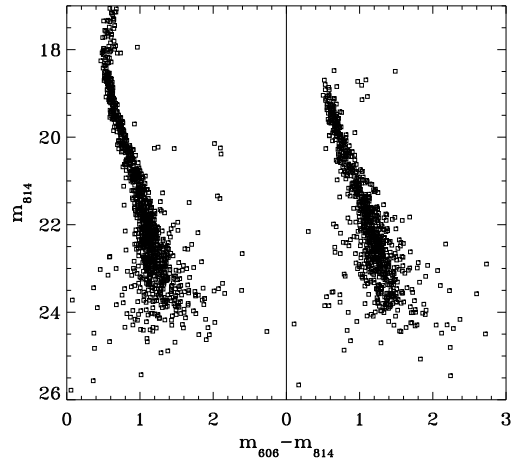}
\caption{ Color-magnitude diagrams of the stars in Field\,1 (left) and
Field\,2 (right). The main sequence in Field\,2 is truncated above
$m_{814}\simeq 18.5$ because of saturation.}
\end{figure}

Instrumental magnitudes have been converted into the WFPC2 ground
system by using the photometric calibration prescriptions for the WFPC2
(Holtzman et al. 1995). These have been used to construct color magnitude
diagrams (CMDs) extending down to $m_{814} \simeq 25$, as shown in
Figure\,2. The internal accuracy of the photometry varies in both
fields from a few hundredths of a magnitude for bright, unsaturated
stars at $m_{814} \simeq 19$ to $\sim 0.1$\,mag at $m_{814}\simeq 24$,
but the photometry in Field\,2 is consistently noisier than in Field\,1
because of crowding, which makes background determination more
uncertain in a denser environment (the number of stars per unit area is
more than three times larger in Field\,2). This effect can be seen by
comparing the ``width'' of the MS in both fields and noticing that the
MS in Field\,1 is narrower than the other at any given magnitude.  The
MS of $\omega$\,Cen is truncated at the bright end because of
saturation, setting in at $m_{814}\simeq 18$ in Field\,1 and
$m_{814}\simeq 19$ in Field\,2 owing to the different exposure duration.

From the CMDs of Figure\,2, and adopting the $2.5\,\sigma$ clipping
criterion described in De~Marchi \& Paresce (1995), I have measured
the LF of MS stars by counting the number of objects in each $0.5$\,mag
bin along the $m_{814}$ axis and within $\pm 2.5$ times the $m_{606} -
m_{814}$ color standard deviation around the MS ridge line. The
contamination due to field stars was removed by counting the number of
objects falling outside of the quoted $\pm 2.5 \sigma$ belt but still
within the $0.5 < m_{606} - m_{814} < 3$ color range. This number has
been scaled by the effective color range ($2.5 - 5\sigma$ mag),
multiplied by the ``width'' of the MS belt (i.e.  $5\sigma$ mag) to
determine the expected number of field stars falling within the cluster
MS, and finally subtracted from the MS star counts.  The selection of
the above color limits stems from the working assumption that field
objects fill almost uniformly the quoted range. The density of objects
on both sides of the MS in Figure\,2 is rather low, in agreement with
theoretical expectations (M\'endez \& Van Altena 1996). It is,
therefore, unlikely that we are underestimating the number of stars in
the MS due to field contamination (see also Elson et al. on this issue).

A reliable determination of the LF from the CMDs of Figure\,2 requires
the assessment of the degree of photometric incompleteness as a
function of the magnitude. I have, therefore, run a number of trials
by adding artificial stars to the individual frames, and then combined
and reduced them again with the same parameters used in the scientific
investigation (see De Marchi \& Paresce 1995 for more details).  In
this way, I have determined that photometry in Field\,1 is $\sim
100\,\%$ complete down to $m_{814}\simeq 22.5$, and that, because of
crowding, the completeness decreases to $\sim 85\,\%$ at $m_{814}\simeq
24$. Field\,2 is more crowded, and photometric completeness is $\sim
100$\,\% for stars brighter than $m_{814}\simeq 20$ dropping to $\sim
50$\,\% at $m_{814}\simeq 24$.

\section{Discussion and Conclusions}

\begin{figure}[h]
\plotone{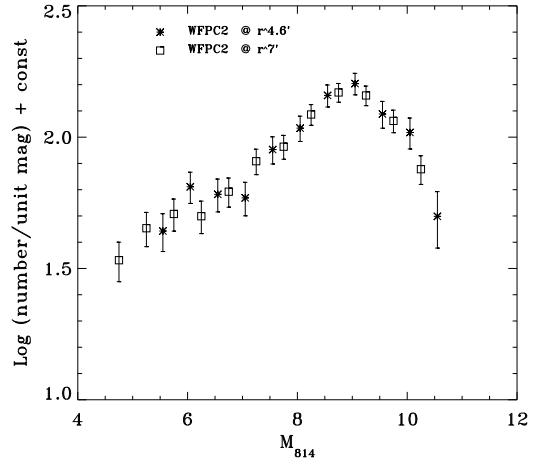}
\caption{Luminosity functions of $\omega$\,Cen measured in Field\,1
(squares) and Field\,2 (crosses) using the CMDs of Figure\,2. The
centers of the bins used to count stars along the MS in the two fields
are displaced by $0.2$\,mag from each other to provide a finer sampling
of the magnitude range.  }
\end{figure}

The LFs measured in this way and corrected for photometric
incompleteness are shown in Figure\,3, where Field\,1 and Field\,2 data
are indicated with boxes and circles, respectively. I have applied a
vertical offset of $0.1$\,dex to the Field\,2 data to account for the
different stellar density in the two images. In addition, I have used
different binning for the data in Field\,1 and Field\,2: the size of
the $m_{814}$ magnitude bin is the same ($0.5$\,mag) for both fields,
but the centers of the bins are displaced by $0.2$\,mag from each
other.  In this way, when the LFs are combined in the same graph
(Figure\,3) they provide a finer sampling of the magnitude range.
Absolute magnitudes in Figure\,3 have been obtained by adopting for
$\omega$\,Cen a distance modulus of $(m-M)_0=13.45$ and a color excess
of $E(B-V)=0.15$ (Djorgovski 1993), thus implying $(m-M)_I=13.7$. The
LFs are in excellent agreement with each other, within the statistical
errors, over a range of more than 5 magnitudes, thus indicating that
both fields are sampling the same stellar population.

The shape of our LFs, however, differs substantially from that derived
by Elson et al., in that those presented here reach a peak at
$M_I\simeq 9$ and then clearly drop all the way to the detection limit,
well before photometric incompleteness becomes significant and
regardless of whether the field at $4\minpoint6$ or the one at
$7^{\prime}$ from the center is considered. This behavior is perfectly
consistent with that observed in all the other low-metallicity GCs
studied so far with the WFPC2 (see references in Introduction). To
confirm this, in Figure\,4 I compare the LFs derived here with the
deepest one today available for a low-metallicity GC, namely that of
NGC\,6397, as measured by Paresce et al. (1995) and arbitrarily
normalized along the vertical axis. In spite of a small difference in
the metal content of the two clusters ([Fe/H]$=-1.59$ for $\omega$\,Cen
and [Fe/H]$=-1.91$ for NGC\,6397; Djorgovski 1993), their LFs agree
very well with each other within the observational uncertainties. This
is not surprising, as both clusters were imaged near their half-mass
radius ($r_{hm} = 4\minpoint8$ for $\omega$\,Cen; Djorgovski 1993),
where the LF should directly reflect the IMF, as the local stellar
population is expected to be rather insensitive to dynamical
modifications (Richer et al. 1991; De Marchi \& Paresce 1997).

\begin{figure}[t]
\plotone{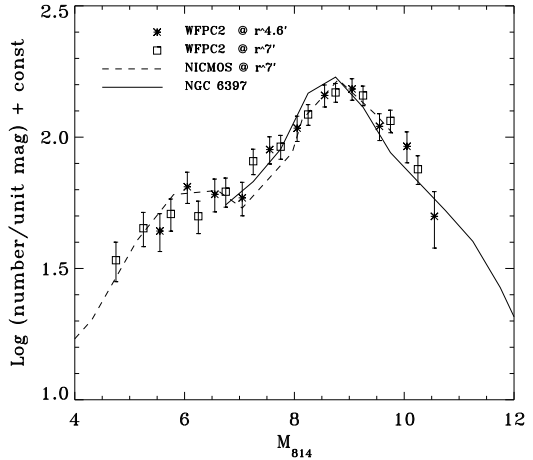}
\caption{The luminosity functions of Figure\,3 are here compared to the
LF of the same cluster measured with NICMOS (dashed line) and converted
to the F555W, F814W plane, as well as with the I-band LF of NGC\,6397. }
\end{figure}

Pulone et al. (1998) have recently observed with NICMOS on board the
{\it HST} a region located $\sim 7^{\prime}$ SW of the cluster center
and have derived a LF for MS stars extending down to $H\simeq 22.5$
($m_{160}\simeq 26$ in the STMAG magnitude system), corresponding to a
mass of $\sim 0.2$\,\Msolar. I have converted this LF from the $H$ to
the $I$ band using the theoretical mass-luminosity relation calculated
by Baraffe et al. (1997) for the metallicity of $\omega$\,Cen, as their
models of stellar atmospheres have shown to reproduce quite accurately
the lower portion of the MS of low-metallicity GCs (Chabrier \& M\'era
1997). The shape of the $I$-band LF obtained in this way (dashed line
in Figure\,4) is in excellent agreement with those measured in the
WFPC2 images over the whole magnitude range spanned by both
instruments. 

It remains difficult to understand the discrepancy between the results
presented here and those published by Elson et al. for their innermost
field, unless the flattening that they measure at low masses is an
artifact due to data processing, and particularly to the lack of
correction for field star and galaxy contamination. On the contrary,
the color information contained in the CMD that I use here should make
the discrimination between cluster and field objects more robust at the
faint end,  where field stars and faint unresolved galaxies begin to
outnumber cluster objects. Although it would be possible to use a model
of the Galaxy to estimate and correct for field star contamination, the
presence of unresolved galaxies and QSOs at $I\simeq 23.5$ and beyond
(see e.g.  Flynn, Gould \& Bahcall 1996) can cause large uncertainties.

In conclusion, my results strengthen the scenario suggested by De
Marchi \& Paresce (1996, 1997) that in order for all the low
metallicity GCs observed so far near the half-mass radius to show very
similar LFs at low masses, they have to be born with the same IMF,
which is now directly reflected in the LF itself with negligible
modifications induced by dynamical evolution.  This proves that it is
possible to determine the IMF of GC stars, provided that a reliable M-L
relation is available and that the LF is measured near the half-mass
radius.

I would like to thank Francesco Paresce for many useful discussions on
the issues addressed in this paper and France Allard and Isabelle
Baraffe for providing their mass--lumi-nosity relations for low-mass
stars.

\end{document}